\documentclass[preprint,12pt]{elsarticle}
\usepackage{xcolor}

\usepackage{amssymb}

\journal{Applied Materials Today}

\begin{document}

\begin{frontmatter}



\author[add1,add2]{Pascal Pochet\corref{mycorrespondingauthor}}
\cortext[mycorrespondingauthor]{Corresponding author}
\ead{pascal.pochet@cea.fr}

\author[add3]{Brian C. McGuigan}
\author[add1,add4]{Johann Coraux}
\author[add3]{Harley T. Johnson}

\address[add1]{Universit\'e Grenoble-Alpes, Grenoble 38000, France}
\address[add2]{Atomistic Simulation Laboratory (L\_Sim) CEA, INAC F-38054, Grenoble, France}
\address[add3]{Department of Mechanical Science and Engineering\\
University of Illinois at Urbana-Champaign, Urbana, Illinois 61801, United States}
\address[add4]{CNRS N\'eel Institut, Grenoble 38000, France}

\title{Toward Moir\'e Engineering in 2D Materials, Using Dislocation Theory}

\begin{abstract}
We present a framework that explains the strong connection in 2D materials between mechanics and electronic structure, via dislocation theory. 
Within this framework, moir\'e patterns created by layered 2D materials may be understood as dislocation arrays, and vice versa.  The dislocations are of a unique type that we describe as van der Waals dislocations, for which we present a complete geometrical description, connected to  both stretch and twist moir\'e patterns.  A simple computational scheme, which reduces the complexity of the electronic interaction between layers in order to make the problem computationally tractable, is introduced to simulate these dislocation arrays, allowing us to predict and explain all of the observed moir\'e patterns in 2D material systems within a unique framework.  We extend this analysis as well to defects in moir\'e patterns, which have been reported recently, and which are the result of defects of the same symmetry in the constituent 2D material layers. Finally, we show that linear defects in the moir\'e space can be viewed as unidimensional topological states, and can be engineered using our framework.
\end{abstract}

\end{frontmatter}


\section{Introduction}
Regular moir\'e patterns appear in many systems in nature, occurring as a result of imperfect stacking of layers.  In this sense, moir\'e patterns may be understood as perfect, periodic networks of defects.  Two dimensional materials provide a platform in which it is possible to find defects in the moir\'e patterns themselves.  The field of 2D materials science has undergone explosive growth in the past decade, with graphene at the leading edge.{\cite{Geim2007}}  Graphene, hexagonal boron nitride (h-BN), and transition metal dichalcogenides (TMDs), for example, are now being used both to discover new fundamental physics, and as building blocks in increasingly sophisticated structures and devices.{\cite{Lee2013}} In applications of these materials, it is usually the case that a 2D material layer is in contact with some other crystalline material layer, and it is this contact that leads to the observation of a moir\'e pattern or features that can be explained by a moir\'e effect.  The simplest examples may be seen in monolayer graphene on crystalline metal substrates,{\cite{VazquezdeParga2008}} or in bilayer graphene (BLG).{\cite{Alden2013}}  

Such interactions may be considered as a form of epitaxy, which has long been associated with film growth of 3D materials, often fairly thick covalent films covalently bonded to similar substrates.  In the 2D materials systems of interest here, however, ``film'' layers may be limited to a single atomic unit cell in thickness.  Furthermore, in 2D materials the interaction between the ``film'' and the ``substrate'' is typically non-covalent in nature, which recalls the notion of van der Waals (VdW) epitaxy, first proposed more than 30 years ago.\cite{Koma1984, Jaegermann2000}

We present here the key components of a framework for moir\'e engineering, or the design of defects in interlayer commensurability with the goal of achieving new system properties.  The building block for the framework is a newly identified class of material defects known as interlayer dislocations,{\cite{Butz2014,Dai2016}} which we term alternatively as van der Waals  dislocations.  In their simplest form, VdW dislocations accommodate crystallographic mis-registry -- either lattice mismatch, rotation, or both -- between ``film'' and ``substrate'', but with interactions that are typically much weaker than in conventional or covalent misfit dislocations.  This gives VdW dislocations some fundamentally different characteristics, such as much larger core radii.  First recognized simply as dislocations in bilayer graphene{\cite{Butz2014}} these defects are in fact ubiquitous in 2D material systems, including materials such as graphene, h-BN, TMDs, molecular crystals, \textit{etc.}, in their interactions with each other and with crystalline solid surfaces such as metal support layers.  Moir\'e beating or interference patterns observed in atomic probe characterization of these interacting systems are simply {\it manifestations} of VdW dislocation networks.  Indeed, we show below that moir\'e patterns -- in principle, {\textit{all}} moir\'e patterns -- can be understood as arrays of dislocations; conversely {\it{all}} periodic arrays of dislocations can be understood as moir\'e patterns. We note here -- for the first time -- that arrays of interface misfit dislocations in conventional strained epitaxial thin film growth, observed in Si$_x$Ge$_{1-x}$ more than 40 years ago, are simply moir\'e patterns with interlayer dislocation core widths that are much smaller than the VdW dislocations observed more recently in 2D materials.{\cite{Kasper1975}}  By recognizing this novel connection to layered structures of 2D materials, and by examining the various classes of possible defects in the moir\'e patterns themselves, we are able to explain numerous observations in the experimental literature that have been seemingly unrelated, including variations in moir\'e pattern intensity in different materials giving rise to nanomesh patterns,\cite{Corso2004,Goriachko2007} the presence of soliton-like defects in BLG systems,{\cite{Alden2013,Kumar2016}} point and line defects in moir\'e patterns,\cite{Tonnoir2013, Lu2014} and  charge density waves in certain 2D TMD systems on graphene.{\cite{Barja2016}}  Finally, we are able to predict new phenomena in layered 2D materials, which together with these other recent observations, may be considered as topological states.  This perspective makes it possible to envision a new field of ``moir\'e engineering,'' beyond the current understanding of the VdW heterostructure concept,\cite{Geim2013} in which moir\'e effects are often only a secondary consideration. 

\section{Moir\'e patterns as arrays of VdW dislocations}

\begin{figure*}[ht]
\includegraphics[width=6.5in]{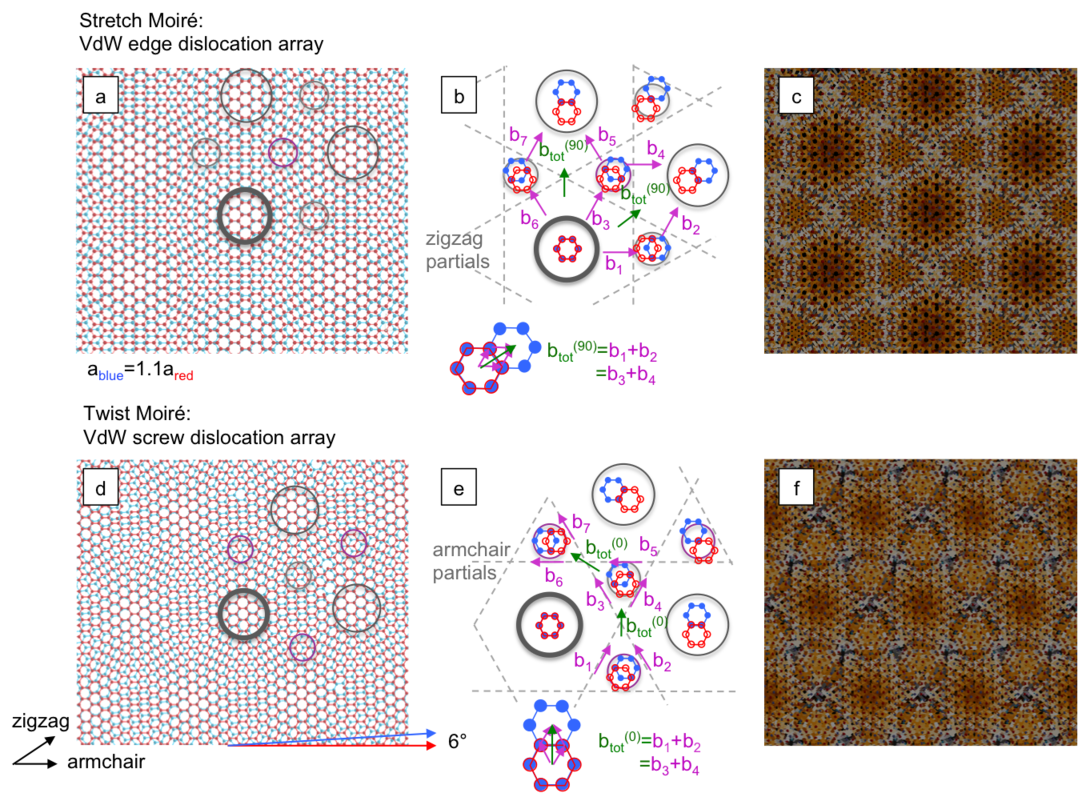}
\centering
\caption{\label{fig:schematic} Van der Waals dislocations are the building blocks of moir\'e engineering.  (a) Red and blue hexagonal lattices with a lattice mismatch strain (here 10$\%$) form a moir\'e pattern.  (b) The moir\'e pattern consists of an array of VdW partial dislocations of \textit{edge} type oriented along the zigzag directions of the lattice, with pairs of partials summing to full dislocations.  The circles here correspond to the circles in part (a); the red lattice is taken as a reference, and the Burgers vectors show the displacement from the red lattice to the blue lattice.  (c) The simulated moir\'e pattern, accounting for van der Waals interactions between the layers, shows the edge VdW partial dislocation array.  The simulation shows a measure of the van der Waals bond density, according to the simple scheme described in Methods.  (d) Red and blue hexagonal lattices with a relative rotation (here 6$^o$) form a moir\'e pattern.  (e) The moir\'e pattern consists of an array of VdW partial dislocations of \textit{screw} type, oriented along the armchair directions of the lattice, with pairs of partials summing to full dislocations.  (f) The simulated moir\'e pattern, accounting for Van der Waals interactions between the layers, show the screw VdW partial dislocation array.}  
\end{figure*}

VdW dislocations can be understood to be lines separating regions of different registry or misalignment between two crystalline layers, as shown in Figure \ref{fig:schematic}.  VdW dislocations have many of the same properties as conventional crystalline dislocations, such as splitting of full VdW dislocations into partials, classification of VdW dislocations as having edge or screw character, and consequences of VdW dislocation motion and pile-up, all of which have implications for 2D material heterostructures.  Arrays of VdW dislocations make up the moir\'e structures observed in scanning tunneling microscopy (STM) and atomic force microscopy (AFM).  While these structures have recently been described as solitons in bilayer systems,\cite{Alden2013} the VdW dislocation terminology more completely describes the atomistic structure of the defects.  Unlike a conventional interface misfit dislocation, which is found at the interface of two lattice mismatched strongly-interacting covalent materials, a VdW dislocation exists between two relatively mismatched or rotated layers, which may be only weakly interacting or even noninteracting.  Thus, the VdW dislocation can have a much more delocalized core structure than a covalent interface misfit dislocation.

Arrays of VdW dislocations created by stretch and twist are shown schematically in Figure {\ref{fig:schematic}} for the representative case of uniform hexagonal lattices; the latter is precisely the same as the bilayer graphene structure.  The dislocation lines themselves are defined in Figure {\ref{fig:schematic}}, and separate regions of so-called ``$H$" stacking, shown in the large circles, from alternating regions of Bernal or ``$T_A$" or ``$T_B$" stacking, shown in small circles.  Based on the classical Peierls-Nabarro model,{\cite{Peierls1940,Nabarro1947}} one supposes that the VdW dislocation has a distributed core width, from one region of perfect stacking to the next, given by, 
 
\begin{equation}
 S(x)=\frac{b}{\pi}\tan^{-1}{\frac{x}{\zeta}}+\frac{b}{2}
\label{corewidth}
\end{equation}

\noindent where $b$ is the Burgers displacement.  The distributed dislocation core half-width is given by $\zeta=K{b}/4\pi\tau^{ideal}$, where $K$ depends upon whether the dislocation is of screw or edge type, and $\tau^{ideal}$ is the ideal shear strength of the interface, which is typically orders of magnitude smaller than in covalent systems.  

Indeed, in the limit of zero force interaction between the two constituent layers, the VdW dislocation core may extend across the entire period of the mismatched unit cell between film and substrate, resulting in a periodic registry across the interface as shown in Figure {\ref{fig:schematic}}a and consistent with an extended core width described in eq. \ref{corewidth}.  It is convenient to assume that the two constituent layers share a perfect (\textit{e.g.} $n/(n+1)$ or $(n+1)/n$) rational relationship between their unstrained lattice parameters, otherwise, in fact, the ``unit cell'' is infinite in size, strictly speaking.  This will result in infinitely many moir\'e orders, a result that can be readily seen through a Fourier space analysis.  The same assumption of perfect rational commensurability is typically made in the analysis of covalent epitaxial systems, resulting in the notion of atomic registry over some periodic unit cell; for example, a system with 2\% lattice mismatch could be modeled by a unit cell containing 50 lattice parameters in one material, and 51 lattice parameters in the other material.  Across one such unit cell, this registry (or anti-registry, equivalently) is the basis for the moir\'e pattern observed in STM or AFM images of graphene, h-BN, and other 2D materials on crystalline support layers.  The resulting moir\'e patterns for the two cases shown in Figure {\ref{fig:schematic}} are shown in parts (c) and (f), for the stretch and twist moir\'e examples, respectively.  These patterns are created using a simple convolution scheme that mimics the electron density shared between the layers as a result of van der Waals interactions.  In this scheme, a measure of the local van der Waals bond density is created by summing up bond contributions across the gap between the interfaces from all bonds within a small range of lengths. It is important to note that this method does not explicitly account for the electronic origins of out-of-plane atom displacements.  But by introducing a variable cut-off radius for the van der Waals interactions, out-of-plane displacements can be considered implicitly, and we make the problem computationally tractable and useful as a design tool for scanning through the numerous material combinations and possible defect structures.  For systems large enough to contain full moir\'e patterns, with defects, first-principles electronic structure calculations are computationally out of reach.  The \textit{bond-convolution} scheme is introduced in the next section and is fully detailed in the methods session.   The result mimics the characterization data from STM and AFM, and shows the distinct nature of the moir\'e pattern for the stretch and twist cases in this limit of zero force interaction between layers.

The geometry of a moir\'e pattern, or superlattice as it may be called, is described in detail in several recent papers{\cite{Cosma2014,Wallbank2015,Hermann2012,Zeller2014,Artaud2016}} for the case of two hexagonal lattices placed rigidly in contact.  This arrangement is shown graphically in Fig. \ref{fig:schematic} (and below, for the case of a hexagonal lattice placed atop a triangular lattice.)  For the simpler case of the pair of hexagonal lattices, the first order moir\'e geometry is described{\cite{Wallbank2015}} in terms of the lowest six reciprocal lattice vectors of the moir\'e pattern, for which ($n=0,...,5$), or 

\begin{equation}
{\vec{b}}_n={\vec{g}}_n-{{\vec{g}}_n}^{\prime}
\label{reciprocal_lattice}
\end{equation}

\noindent which may be written as ${\vec{b}}_n \approx \delta{\vec{g}}_n-\theta{\vec{l}}_z \times {\vec{g}}_n$, since ${{\vec{g}}_n}^{\prime}$ is the ``substrate'' reciprocal lattice vector, with small stretch $\delta$ and small rotation angle $\theta$ relative to the ``film'' hexagonal lattice.  The real space lattice vectors of the moir\'e pattern are therefore readily computed by inverting eq. \ref{reciprocal_lattice}.


While the moir\'e pattern may therefore be represented as points in a lattice, we focus on a representation relying on the VdW dislocation array.  The VdW dislocation lines in Figure {\ref{fig:schematic}}, shown as dashed grey lines in parts (b) and (e), have Burgers vectors given by $\bf{b}_i$, where the direction of the Burgers vector denotes not only the direction of slip between neighboring regions of registry, but also the screw or edge character of the VdW dislocation.  For example, in the case of two hexagonal lattices with a relative lattice mismatch but no relative rotation, shown in Figure {\ref{fig:schematic}}a, the Burgers vector is perpendicular to the local line direction $\xi$, and the VdW dislocation array is thus entirely of edge type, as shown in part (b): the partials sum to form 90 degree or edge-type full dislocations.  
Furthermore, the VdW dislocation array is of \textit{edge} type and partial dislocations are aligned along the zigzag direction of the constituent lattices.  When there is a twist angle between the layers, as in Figure {\ref{fig:schematic}}d, the Burgers vector $\bf{b}$ is entirely parallel to the local line direction $\xi$, and thus the VdW dislocation is of \textit{screw} type; again, the Burgers vector corresponds to the direction of relative slip between regions of neighboring registry.  The partials sum to form zero degree, or screw-type full dislocations.  The VdW partial dislocations are, furthermore, aligned with the armchair direction of the constituent lattices.  It is worth noting that moir\'e patterns in the stretched and twisted cases are almost identical and differ only by their orientation with respect to the reference atomistic lattice: 30$^{\circ}$ rotated (in the stretch case) but aligned (in the twist case).  This similarity might explain why the different \textit{edge}- and  \textit{screw}-type VdW partial dislocation types are easily overlooked.

The simplest examples of these VdW partial dislocation networks occur in bilayer graphene where there is often a relative rotation between the layers, similar to a case observed by Alden et al.\cite{Alden2013}  The observed moir\'e structure in such cases will vary slightly depending on the strength of the van der Waals interaction between layers and the relative rotation angle, which affects the dislocation core width, but the symmetry of the pattern will be fixed by the dislocation geometry illustrated in Figure \ref{fig:schematic}e.  In this case, the VdW dislocation array is of \textit{screw} type, and the features of the moir\'e pattern, sometimes referred to as solitons, are a manifestation of that VdW dislocation array.  Indeed the former study \cite{Alden2013} may be one of the first experimental studies to explore VdW dislocation motion, which is a promising area of inquiry, but beyond the scope of the present analysis.  VdW dislocations may also have mixed screw and edge character, as in the case of bilayer h-BN-graphene where the different B-C and B-N chemistry naturally adds an edge character (due to stretch) when compared with the pure screw bilayer-graphene structure (due to rotation). It is worth noting that in this case the amount of screw character ({\it{i.e.}} the angle between the two layers) has a strong influence on the resulting properties, and that the strength of the van der Waals interaction can influence the resulting screw character of the VdW dislocation array.\cite{Tang2013}

\begin{figure*}[ht]
\centering
\includegraphics[width=6.5in]{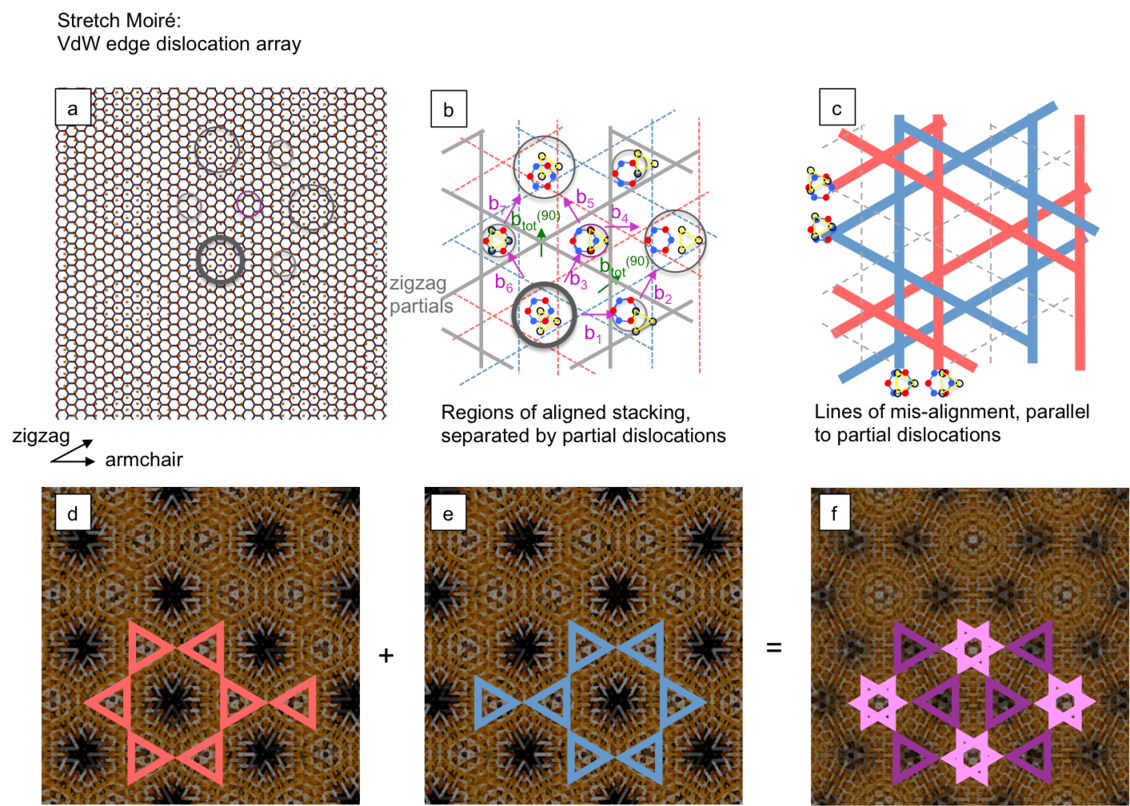}
\caption{\label{fig:misalignment} VdW partial dislocations separate regions of crystallographic alignment between layers; moir\'e patterns are created by misalignment.  (a) A six-fold symmetric lattice (\textit{e.g.} graphene) on a three-fold symmetric lattice (\textit{e.g.} Ru(0001). Regions of near-perfect stacking are circled.  (b) In the regions of near-perfect stacking, the relationship between the hexagonal lattice and the triangular lattice is illustrated, where the hexagonal lattice is understood to be the reference, and the triangular lattice shifts according to the burgers vectors.  The partial dislocations (grey solid lines) represented in the moir\'e pattern separate three types of stacking in the case of the hexagonal lattice on a triangular lattice.  These Shockley partials are of edge type, and sum according to the simple rules shown.  (c) The wide cores of the partial dislocations are bounded by lines of mis-alignment between atoms in the two layers, shown as red and blue lines that are parallel to the partials; these lines indicate directions along which the triangular lattice atoms are slightly mis-aligned relative to the red atoms or blue atoms in the hexagonal lattice. (d,e,f) These lines of misalignment form networks of triangles that sum to create the moir\'e pattern; the triangles can be considered as a simple ball-and-stick model for the moir\'e pattern.  The moir\'e pattern in this case is the sum of two honeycomb or nano-mesh patterns, each of which is representative of the moir\'e associated with a six-fold symmetric lattice with unequal sublattices (e.g. h-BN) on a three-fold symmetric lattice (e.g. Ru). 
} 
\end{figure*}

In addition, the concept of the VdW dislocation can be applied to cases in which hexagonal structures such as graphene or h-BN layer align with lattices of triangular symmetry, such as dense-packed metal surfaces (Ir, Co, Ru, etc.) on which they are commonly grown.  Indeed, moir\'e patterns are readily observed in these cases, and can be shown to be arrays of VdW dislocations as depicted in Figure {\ref{fig:misalignment}}a, where a hexagonal lattice, in this case with diatomic composition, is placed in contact with a triangular lattice that has a relative stretch, so the lattice parameters are mismatched.  Such a configuration would be found in the case of h-BN on a Ru(0001) substrate.   Figure {\ref{fig:misalignment}}b shows the patterns of alignment (in circles) that are separated by arrays of VdW partial dislocations (grey lines); these circles are also shown in Figure \ref{fig:misalignment}a.  In this case, the three ``perfect'' stacking types are ``$T$'' (large circles) and alternating regions of ``$H_A$'' or ``$H_B$'' (small circles).  The regions are separated by a VdW partial dislocation network with the same symmetry as that of the hexagonal/hexagonal case shown in Figure {\ref{fig:schematic}}b.  The Burgers vectors, again, show the displacement of the triangular lattice relative to the hexagonal lattice as one moves between these regions of perfect stacking.  The partial dislocations, shown as grey lines, evidently Shockley partials, sum to form 90 degree or edge type full dislocations.

In Figure {\ref{fig:misalignment}}c we introduce a new concept to explain the connection between the VdW partial dislocation network and the resulting moir\'e pattern.  Here we identify the lines of maximum misalignment (red and blue) between the A and B sublattices of the hexagonal lattice (red and blue dots), and the triangular lattice (yellow dots).  The atomic alignment is sketched alongside the lines of misalignment.  This construction shows clearly that there are two families of such lines in the bilayer system -- one corresponding to each family of partial dislocations (that originate, respectively, from the two sublattices in the hexagonal lattice).  These lines of misalignment are the conceptual connection between the VdW dislocation structure and the resulting moir\'e pattern.  The details of this relationship are more complicated in the case of nonzero van der Waals interactions, such as in the bilayer h-BN-graphene system.\cite{vanWijk2014}

In Figure {\ref{fig:misalignment}}d and  {\ref{fig:misalignment}}e we show that these lines of misalignment are equivalently but more clearly represented as networks of triangular regions.  In this form, the objects still represent maximum misalignment between the two respective (red and blue) sublattices in the hexagonal lattice relative to the underlying triangular lattice.  As such, in a system with unbroken symmetry between these networks, the triangles constitute perfect, complementary networks that sum to form the moir\'e pattern.  Each network has the appearance of honeycomb symmetry, or a ``nanomesh'' as seen experimentally \cite{Corso2004,Goriachko2007}, and shown in Figure {\ref{fig:misalignment}}d and  {\ref{fig:misalignment}}e.  When superimposed, the networks give a triangular lattice of circles, or mounds; thus, Figure {\ref{fig:misalignment}}d and Figure {\ref{fig:misalignment}}e sum to give Figure {\ref{fig:misalignment}}f.  The triangle representation is analogous to the ball-and-stick representation commonly used to understand symmetry in atomic $lattices$; the lines of misalignment are more appropriately considered to be a $network$ than a lattice of points.  The simulated moir\'e pattern for the complete system is shown in Figure {\ref{fig:misalignment}}f.  Again: this pattern is the sum of moir\'e patterns associated with the individual A and B sublattices interacting with triangular lattice, shown in Figure {\ref{fig:misalignment}}d and {\ref{fig:misalignment}}e, respectively.  

The construction shown here is useful both in understanding the relationship between the VdW partial dislocation array and the moir\'e pattern in a hexagonal lattice on a triangular substrate and for understanding the difference between graphene and h-BN moir\'e patterns on triangular substrates.  In the graphene case one assumes that both families of VdW partial dislocations, and the misalignment patterns that accompany them, are equally active.  Thus, graphene on Ru(0001) results in a triangular moir\'e pattern like that seen in Figure {\ref{fig:misalignment}}f.  In h-BN on Ru(0001), one assumes that either the B or the N sublattice interacts more strongly with the substrate, thus breaking the symmetry.  The resulting moir\'e pattern is thus more like either Figure {\ref{fig:misalignment}}d or {\ref{fig:misalignment}}e.  This difference is explored in more detail below.  
Interestingly, the same degeneracy lifting has been observed between graphene and some Co,{\cite{Decker2013} which we hypothesize to be due to an interplay between the graphene and the 2nd neighbors in the metal substrate, which breaks the symmetry between the two VdW partial dislocation sublattices. Such a symmetry breaking effect may also be induced by some localized doping (\textit{e.g.} a simple vacancy), and may  propagate over a large  area.\cite{Lawlor2014}

\section{Moir\'e defects revisited through the VdW dislocation framework}

Thus far, we have shown that VdW dislocations, including their partial and sub-lattice character,  reproduce the different observed contrast in moir\'e patterns.  Conceptually, the VdW dislocation array may be understood as an electronic structure feature of the system in that it exists due to van der Waals interactions in the interface between the film and substrate.  This has profound implications for  basic physics that is manifested, as well as for the potential device applications of the moir\'e patterns.  By understanding that the network properties arise from the basic science of dislocations, it becomes clear that it may be possible to engineer systems by combining various layers, with or without underlying covalent defects, in order to exploit a variety of interesting outcomes. Moir\'e defects are readily observed experimentally, typically using scanning tunneling microscopy, where the measurement probes the tunneling current due to the local density of states in the combination of constituent layers.  Thus, to study these outcomes it is useful to generate simulated moir\'e images in various levels of approximation.  We note here that the forces between the layers are typically weak for the systems of interest, compared with the forces binding the atoms within the film or within the substrate.  The moir\'e pattern, however, is a manifestation of electronic structure interactions in the interface controlled by the strong in-plane structure.  In this sense the moir\'e pattern, and any defects it may contain, are topological states.  

For the purposes of simulating the moir\'e and demonstrating the existence and properties of VdW dislocations, we avoid using a highly expensive computational method such as DFT, for two reasons: it is challenging to accurately parameterize van der Waals interactions using first-principles methods, and, we note again, the calculations are very costly for systems sufficiently large to simulate VdW dislocation networks of any practical size.  Initial efforts to model simple moir\'e patterns using electronic structure based methods have been successful, but only for perfect systems. {\cite{Brugger2009, Kang2013}}  Methods are only now emerging for more accurate representation of force interactions at sufficiently large scales that defects may be present; still, these methods do not directly simulate electronic structure.{\cite{Zhang2016}}  Thus, we adopt a simple model that renders simulated images based on a plot of van der Waals bond density; bonds that have lengths within a predefined range are included in constructing the image, mimicking interactions between film and substrate atoms. In this way, the model \textit{implicitly} considers variation in local VdW bond density -- in the important limit of weak VdW interactions -- through dependence on local bond length.  As in STM images of these systems, the computational model also discriminates between pair-wise interactions between different atomic species, which reflects differences in both chemistry and strain and leads to locally different moir\'e intensity.{\cite{vanWijk2014}}  This bond superposition technique is used to generate a representation of multilayer stacking that more directly represents the local electron density between layers using a real-space analysis.  The scheme requires a simple procedure that is summarized in the Methods section below.  The simulation approach results in Figures {\ref{fig:schematic}}c, {\ref{fig:schematic}f and {\ref{fig:misalignment}}d-f, and makes it possible to reproduce an astonishing variety of perfect VdW dislocation arrays.  

\begin{figure*}[ht]
\centering
\includegraphics[width=6.0in]{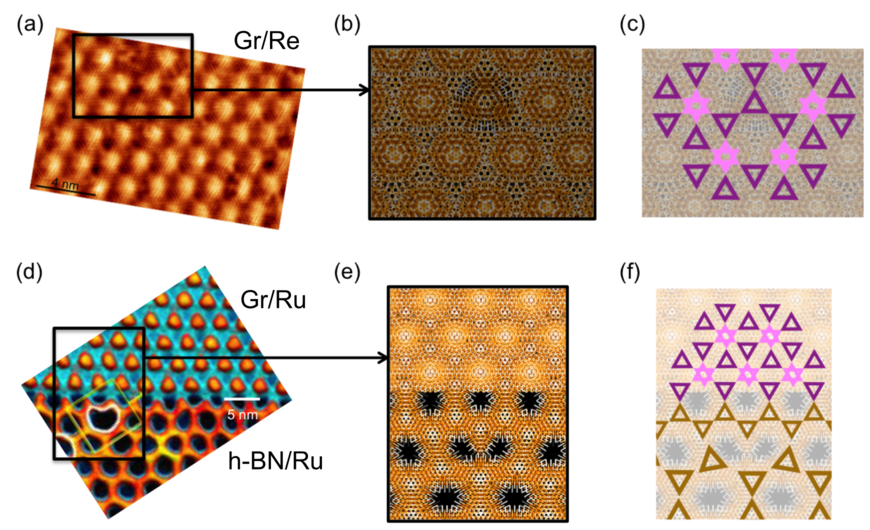}
\caption{\label{fig:Moire_disloc} Defects in the moir\'e pattern can be explained by broken symmetry in the pattern of misalignment.  (a) The experimental moir\'e pattern associated with graphene on Re contains occasional ``vacancies". \cite{Tonnoir2013,Qi2017} (b) The simulated moir\'e pattern contains an identical vacancy defect when one of the families of van der Waals partial dislocations is missing in a localized region.  (c) The ``ball-and-stick" representation of the same defect shows how a missing region of misalignment gives rise to the defect.  (d) The graphene/h-BN interface, when on a Ru support layer, features occasional eight-sided ring moir\'e defects {\cite{Lu2014,Qi2017}} that are signatures of underlying 5-7 or 8-6 atomistic interface misfit dislocations \cite{McGuigan2016}.  (d) The simulated moir\'e pattern, when the atomistic lattice contains such a defect, shows a eight-sided ring moir\'e defect near the interface, with contrast on the graphene and h-BN sides of the interface matching that of the experimental image.  (e) The symmetry of the moir\'e defect can be understood by constructing the lines of mis-alignment that occur between the graphene/h-BN heterostructure and the Ru substrate, as described in Figure 2.  The moir\'e interface in this case can be said to be incoherent, while in the absence of the strain-relieving atomistic 5-7 or 8-6 dislocation, the moir\'e interface would remain coherent (as shown in the Appendix ).} 
\end{figure*}

\section{Localized Defects in Moir\'e Patterns.}

Like any periodic array, VdW dislocation arrays themselves may contain defects, which appear as breaking of symmetry in the corresponding moir\'e patterns.  Figure \ref{fig:Moire_disloc} shows two examples of moir\'e patterns with localized defects, from the recent experimental literature of \textit{sp$^2$} materials on HCP metal,\cite{Tonnoir2013,Lu2014} and from the results of the VdW dislocation-based model.  The moir\'e defects shown here are both due to localized features in the \textit{sp$^2$} material, but are better understood by analogy to bulk materials, where one is more similar to a classic point defect, and the other to a classic line defect or dislocation. 

Figure \ref{fig:Moire_disloc}a shows an experimental STM image of example of a localized or point defect in the moir\'e pattern observed in the case of graphene on Re.\cite{Tonnoir2013}  This defect has been studied more recently in h-BN/graphene lateral heterostructure growth.\cite{Qi2017}  The image shows clear evidence of a missing moir\'e spot, or a vacancy.  Similar defects are widely observed in the case of CVD graphene on a variety of metal substrates, but interestingly, they are more commonly seen for metal substrates with which graphene is known to interact strongly.  Viewed within the context of the VdW dislocation array, it is obvious that such a moir\'e defect is due to a local change in registry between the graphene and the metal substrate, thus eliminating one of the spots that indicates a particular region of registry; the simulated result is shown in Figure {\ref{fig:Moire_disloc}}b.  This defect is analogous to cases in which a covalent interface misfit dislocation reconstructs to annihilate a junction, which is widely seen in TEM images of relaxed SiGe/Si.{\cite{Kasper1975}}  In the graphene case, however, we can explain the appearance of this moir\'e defect using the misalignment ball-and-stick model introduced in Figure {\ref{fig:misalignment}}, as shown here in Figure {\ref{fig:Moire_disloc}}c.  Within our framework, the moir\'e vacancy can be understood as the result of a local absence of one of the two families of VdW partial dislocations, and thus an absence of the corresponding lines (or triangles) of misalignment.  The vacancy is not entirely without STM contrast, as there is misalignment locally, but the total contrast is significantly reduced.  One may similarly postulate that other such point defects, perhaps even interstitial defects, may exist under the right conditions.  The details of the atomistic structure in the graphene layer, or the Re layer, are not known, and may in fact not be unique. 

A VdW dislocation network also admits defects analogous to edge dislocations, as seen recently in several experiments in h-BN/graphene lateral heterostructures on metal support layers.\cite{Lu2014, Sutter2012, Qi2017}  In these experiments, a monolayer graphene/h-BN heterointerface is prepared, with the structure supported on a crystalline metal surface, \textit{e.g.} Ru(0001).  The moir\'e pattern contains 1D defects for which the line direction can be defined, into the plane of the 2D layer, like the line direction $\xi$ of a crystalline dislocation.  As in the case of a crystalline dislocation, the defect may be understood as an interfacial misfit dislocation, and the result of the in-plane interface, as shown in Figure \ref{fig:Moire_disloc}.  This feature is not precisely analogous to an edge dislocation in an atomic lattice, because the moir\'e is better described in this framework as a network of lines and not as a lattice of points, but it shares many of the same features as an edge dislocation in a covalent network, including the equivalent to a Burgers vector component perpendicular to the line direction.  Furthermore, and most interestingly, it can be readily seen that the defect is {\it{dual}} to the underlying covalent defect that gives rise to it.  Here, an 8-6 dislocation in graphene is at the core of the defect in the moir\'e (shown in Supporting Information) and is a result of strain relaxation in the graphene/h-BN interface.{\cite{McGuigan2016}}  The identical 8-6 defect symmetry can be seen at the moir\'e scale, in agreement with the recently reported moir\'e  magnification of defects. \cite{Wallbank2015}

This magnification effect is well reproduced within the VdW dislocation framework as shown in Figure \ref{fig:Moire_disloc}e and in Figure \ref{fig:interface_atomistic}.  Indeed, the simulated moir\'e pattern, which agrees very well with the experimental image, including both the symmetry of the moir\'e defect and the distinct moir\'e structures seen on the graphene/Ru side of the in-plane interface when compared with the h-BN/Ru side.  The difference between the h-BN and graphene regions is explained in Figure \ref{fig:misalignment}.  The incoherent interface in the present case is readily explained through the ball-and-stick construction shown in Figure \ref{fig:Moire_disloc}f.  Interestingly, this moir\'e interface reconstruction is not unique: it is sensitive to small translations of the graphene/h-BN layer on the Ru substrate.  Furthermore, one finds a coherent moir\'e interface over regions of the graphene/h-BN covalent interface where there are no underlying interface misfit dislocations, such as in the region outside the inset box shown in Figure \ref{fig:Moire_disloc}d (and as shown in Figure \ref{fig:coherent_supp})

\begin{figure*}[ht!]
\centering
\includegraphics[width=6.5in]{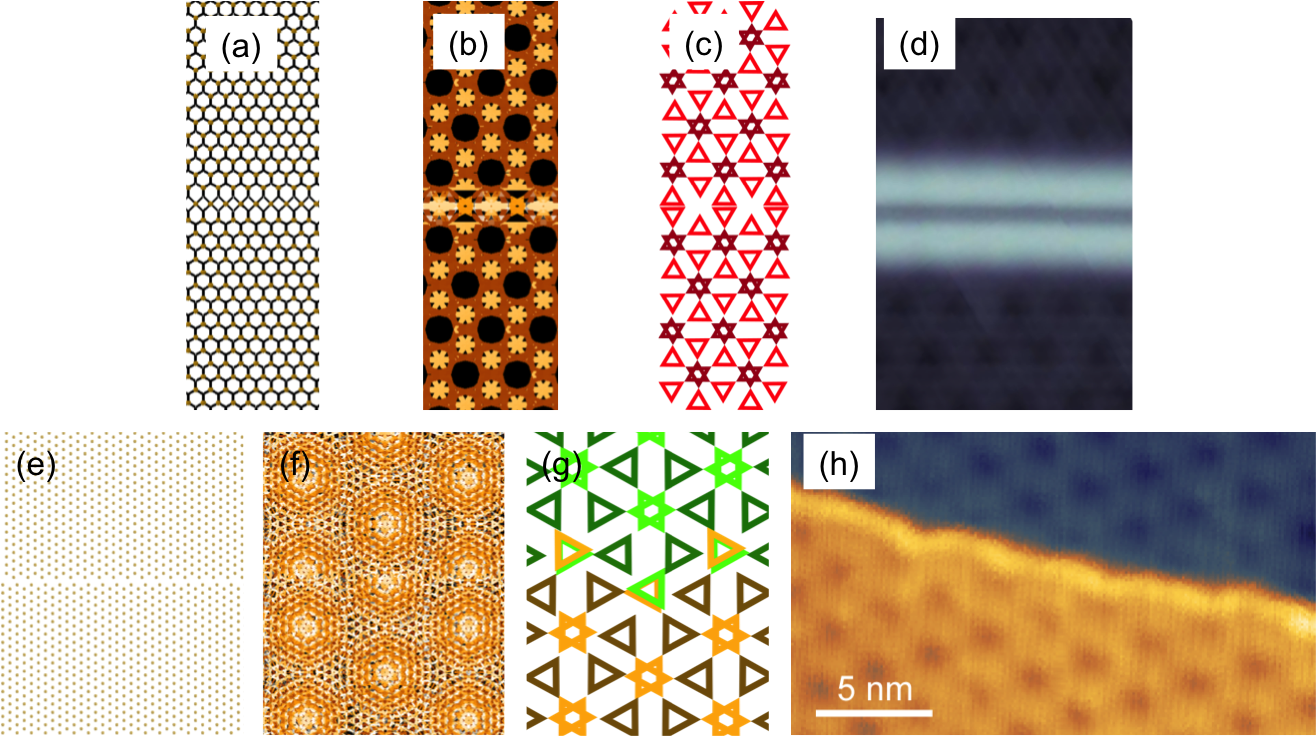}
\caption{\label{fig:Moire_line_defects} Moir\'e line defects are associated with line defects in the covalent lattice.  (a) A ``mirror twin" anti-phase boundary in an MoSe$_2$ layer.  (b)  When this layer is supported on graphene, a topological state is seen along the boundary; the moir\'e pattern is more coarse than in the graphene/Ru system, because the mismatch between MoSe$_2$ and graphene is larger.  (c)  The pattern of misalignment produces a moir\'e interface reconstruction.  (d) A charge density wave is observed along such a mirror twin boundary in the MoSe$_2$/graphene system, by Barja et al.{\cite{Barja2016}} (e) A screw-type anti-phase boundary on an Ir(111) surface due to a small shift between two grains, along the $<$-211$>$ direction.  (f) When a graphene monolayer is placed on this surface, with armchair direction along the boundary, a moir\'e pattern with a shift along the boundary appears as a sinusoidal topological state.  (g)  This topological state can be explained by the pattern of misalignment in the ball-and-stick representation.  (h)  The scanning tunneling microscope image of such a system, shows a state corresponding to enhancement of the local density of states along the boundary.  An in-plane oscillation is seen and is found to be commensurate with the repeat distance between the highest symmetry moir\'e direction (see the direct comparison in Figure \ref{fig:Johann_exp_large}).} 
\end{figure*}
 
\section{Toward Moir\'e engineering}   
 
The two moir\'e defects presented in Figure \ref{fig:Moire_disloc} derive from the growth of the \textit{sp$^2$} material on a metal and are the unavoidable result of energy minimization in the 2D material layer.\cite{McGuigan2016}  Their moir\'e structure is well reproduced by our VdW dislocation framework. Next, we use this framework to predict systems with possible topological states in the moir\'e space.  Two-dimensional or planar defects, which appear as boundaries in moir\'e patterns, are ideal candidates for this purpose.  The resulting moir\'e defects are not simply lines in a regular moir\'e pattern, but are instead defects due to grain boundaries (e.g. stacking faults, twins, antiphase boundaries, low angle grain boundaries, etc.) in the underlying covalent lattice(s).  The features can be described as topological states and may display unique electronic transport behavior.{\cite{Jaskolski2016}}  One concludes through the examples shown below, for example, that an antiphase boundary (APB) in the covalent network should lead to related antiphase boundary defect in the moir\'e pattern. 

In Figure \ref{fig:Moire_line_defects}a and b we present such an antiphase boundary in MoSe$_2$ and the resulting moir\'e pattern when combined with graphene.  It is worth noting that the scale of the moir\'e pattern is reduced compared with the examples shown in Figure \ref{fig:Moire_disloc} (and thus the resolution in the moir\'e contrast is more coarse) due to a greater mismatch between the two layers.  This APB moir\'e pattern is sketched with the ball-and-stick representation in Figure \ref{fig:Moire_line_defects}c.  In the duality between the covalent and the moir\'e APB structures, both correspond to a perpendicular relative displacement of two domains along the boundary, which may be described therefore as an \textit{edge}-type antiphase boundary. Interestingly a closely related APB has been observed recently in the MoSe$_2$/graphene system by Barja et al\cite{Barja2016} at both the covalent and the moir\'e scales, the latter of which is shown in Figure \ref{fig:Moire_line_defects}d.  While the cases in Figure \ref{fig:Moire_line_defects}a-c consider an infinite APB, the experimental image in Figure \ref{fig:Moire_line_defects} corresponds to a few-nm edge of a triangular domain.  Despite this difference, the simulation accurately captures the main features of the moir\'e pattern, although the enhanced charge density is not reproduced as well. (A larger-scale comparison is shown in Figure \ref{fig:Barja_supp}).  In particular, the symmetries are well reproduced;  the highest intensity lines in the moir\'e APB are well apart from the covalent APB and overlap the innermost lines of moir\'e spots in each domain.  The same lattice parameter is observed in the model and in the experiment.  Based on observations from our framework, we raise the possibility that the charge density wave (CDW) in the recent experiment is not only  a property of the MoSe$_2$ layer but also related to the film/substrate van der Waals interactions.  In other words, based on the moir\'e simulation, freestanding MoSe$_2$  may not be expected to present the same CDW feature that we identify to be connected to this moir\'e APB.  Finally, with respect to this defect, we note that the presence of an additional displacement parallel to the  moir\'e APB in the experiments corresponds to a small \textit{screw}-type component that derives from the VdW interaction between the two layers. This is not considered in our simple simulation (which applies in the limit of zero force interactions).  Such a displacement is made necessary by the particular closure rules imposed by the triangular domains seen in the experiment -- for which the two types of covalent APBs have been identified.\cite{Lin2015}  Indeed, the screw component of displacement along the APB may be necessary for the appearance of a strong charge density signature.

As such, one can expect that other systems should exhibit a similar or stronger signature of the APB. In the first example (shown in Figure \ref{fig:Moire_line_defects}a-d), the edge-type of the APB leads to a natural separation of the edge state of the border of each grain.  Now, we consider a possible protected state that would arise without the need for two grains to be separated in this way.  We seek a pure screw-type APB, a situation which for instance occurs when the lattice of the substrate is shifted in the plane of the surface, and the two shifted domains have a boundary parallel to the shift.  This is precisely the situation encountered at the dense-packed surface of metals, when crossing an atomic step edge.  In the case of a face-centered cubic (FCC) metal cut in the (111) plane, the lattice of one terrace and its neighbor, separated by an atomic step edge, are indeed laterally shifted, by $\frac{1}{6}$(-1 $\frac{1}{2}$ $\frac{1}{2}$), when the step edge is oriented along a $<$-211$>$ direction, according to the FCC stacking.  Such an example is shown in Figure \ref{fig:Moire_line_defects}e for the case of an Ir(111) surface.  Figure \ref{fig:Moire_line_defects}f shows the computed moir\'e pattern for monolayer graphene on this substrate, and Figure \ref{fig:Moire_line_defects}g shows the ball-and-stick image representing the misregistry that leads to this moir\'e pattern.  Scanning tunnelling microscopy of graphene prepared by chemical vapor deposition on Ir(111) reveals indeed that across such a step edge, the expected screw-type APB occurs in the moir\'e, as shown in Figure \ref{fig:Moire_line_defects}h.  Another view comparing the predicted and experimental images is shown in Supporting Information.  Previous works \cite{Coraux2008,Donnor2009} reported edge-type APB in graphene on metals, found at the vicinity of metal step edges in the $<110>$ direction (or $<100>$ in the case of a hexagonal compact metal like Ru).  In Figure \ref{fig:Moire_line_defects}h, we note that the local density of states is enhanced at the location of the step edge, which could point to the presence of an electronic state with limited extension perpendicular to the step edge. The local density of states enhancement shows a spatial undulation in the surface plane, which we find is in agreement with the predicted moir\'e lattice, and motivated by the complete framework described here.

In summary, we show that stacking of 2D material layers on other 2D materials or crystalline substrates, with either hexagonal or triangular symmetry, naturally gives rise to van der Waals dislocations arrays, which appear as moir\'e patterns.  A convenient geometric terminology can be adopted to describe the partial dislocation structure of the VdW dislocation arrays.  Furthermore, a framework for describing the geometry of misalignment gives rise to a simple representation, akin to a ball-and-stick model, which makes it possible to predict moir\'e patterns in good agreement with simulations and recent experimental observations.  Moir\'e patterns are found to contain defects with a precise geometrical dual relationship to defects in the underlying 2D material layers.  This framework helps to explain the existence of novel topological states that may have unique electronic properties, and it permits the prediction of undiscovered moir\'e features across a variety of 2D materials.

The framework leads naturally to the idea that by taking advantage of defects in the constituent layers, it is possible to design topological states in the moir\'e pattern.  In particular, we consider two 1D topological defects that result from antiphase boundaries in one of the two constituent layers.  One example, a linear charge density wave, has been reported recently in the MoSe$_2$/graphene system; the other is a newly discovered sinusoidal charge density signature that appears in the graphene/Ir(111) system.  Furthermore, it can be shown that defects interacting with another layer through van der Waals interactions follow a \textit{commutative} property with respect to the creation of a moir\'e defect (as shown in Figure \ref{fig:zig-zig-defects}); a defect in one layer creates the same moir\'e defect that the equivalent defect in the other layer creates.  This provides another degree of freedom for designing states in stacked 2D material systems, given the ability to engineer defects in the constituent layers.  The idea that it may be possible to engineer moir\'e patterns and their defects is emerging from a variety of experimental observations; it has even been shown recently that by tuning interlayer interactions and controlling what we refer to as VdW dislocation arrays, it is possible to access a new \textit{phase space} of moir\'e patterns.{\cite{Martinez-Galera2016}}  Together with these observations, the framework presented here opens the door to the possibility of engineering van der Waals heterostructures, or moir\'e engineering.

\subsection*{Van der Waals bond density computational scheme}
The simulation of moir\'e images is done via a superposition scheme that creates a representation of van der Waals bond density.  All simulated moir\'e results are assumed to be in the limit of weak VdW force interactions between the two constituent layers. The potential energy surface \cite{vanWijk2014} relating the two layers is considered indirectly, instead, via the choice of cut-off radii considered for each pair-wise atom interaction (e.g. B-Ru, N-Ru, or C-Ru).  This mimics the STM signature, which includes the effects of both chemistry and local strain. Thus, the moir\'e simulation method does discriminate between local atomistic composition, as follows.
\begin{enumerate}
\item A range of cutoff radii $[r_{min},r_{max}]$ is set to control the range of the considered van der Waals attraction between top and bottom layer atoms.  Atoms that fall within this range are assumed to contribute to the VdW interaction.
\item  Contributions to the local electron density are represented visually by line segments connecting pairs of van der Waals interacting atoms, across the gap between the two interacting layers.
\item The cutoff radii differ by species, e.g. in an h-BN/graphene layer on Ru, three different sets of cutoff radii are applied: B-Ru, N-Ru, and C-Ru.  In this way, it is possible to mimic the highly nonuniform composition-dependent interactions that have been reported.\cite{vanWijk2014}  In the analyses presented here, which appear in all four figures, and the figures in the Supporting Information, those values are chosen for each system to exclude the near-neighbor interactions, but to include the second-neighbor interactions between the triangular lattice and the two nearby hexagonal sub-lattices.
\item The total moir\'e field intensity, which simulates the moir\'e pattern as measured by STM, for example, is then expressed as the visual superposition of the line segments between interacting atoms, or VdW bonds.
\end{enumerate}
This method creates a simulated field that mimics the spatially varying Van der Waals interaction between layers, a problem which remains prohibitively computationally expensive for a more accurate electronic structure method such as density functional theory, for moir\'e patterns on the scale of those considered here.  We emphasize that while this does not account for local relaxation of atoms either in the plane of the stacked structure or out of plane, the variable cut-off radius by atomic species provides sufficient variability as to capture the effect of weak force interactions between layers.  The resulting simulated moir\'e patterns are qualitatively very similar to scanning tunneling microscope images of the charge density, or van der Waals interaction, between layers.

\subsection*{Graphene on Ir(111) and Scanning Tunneling Microscopy}
Graphene was prepared by chemical vapor deposition on an Ir(111) surface with ethylene as a carbon precursor. To perform growth, a $5x10^{-9}$ mbar ethylene partial pressure was introduced in the ultra-high vacuum chamber and the sample temperature was held at 1050$ ^\circ$C. Prior to graphene growth, the Ir(111) surface was prepared by repeated cycles of Ar ion bombardment and high temperature flashes. The STM measurements were performed in constant current (30 nA) mode at 130 mV sample-to-tip bias.



\section{Acknowledgements}
H.T.J. acknowledges support from a Fulbright Fellowship and from Universit\'e Joseph Fourier, Grenoble, through an invited professorship.  B.C.M. acknowledges the support of the National Science Foundation Graduate Research Fellowship Program under Grant Number DGE-1144245 and a NSF GROW (Graduate Research Opportunities Worldwide) Award. P.P. and J.C. acknowledge partial funding from French National Research Agency (ANR) through the 2D TRANSFORMER project. Some DFT calculations were performed using French supercomputers GENCI-CINES and GENCI-CCRT through project 6194. We thank many colleagues for helpful technical discussions, including Prof. E. Ertekin, Prof. A. Van der Zande, Prof. T. Michely, Prof. M. Cot\'e, Dr. D. Caliste, Dr. Y. Dappe and  Dr. H. Okuno.

\appendix
\section{}
When an incoherent moir\'e interface is seen, as in Figure \ref{fig:Moire_disloc}, it is result of an atomistic dislocation in one of the constituent crystalline layers.  The example in Figure \ref{fig:Moire_disloc} is due to an 8-6 dislocation on the h-BN side of the h-BN/graphene lateral heterostructure, as shown in Figure \ref{fig:interface_atomistic}.  As observed in Lu et al, 2014,\cite{Lu2014} the defect need not be precisely at the atomistic interface in order to be a strain-relieving feature of the lattice.  But defect in the moir\'e pattern has precisely the same symmetry as the defect in the h-BN.

\begin{figure*}[h]
\includegraphics[width=5.0in]{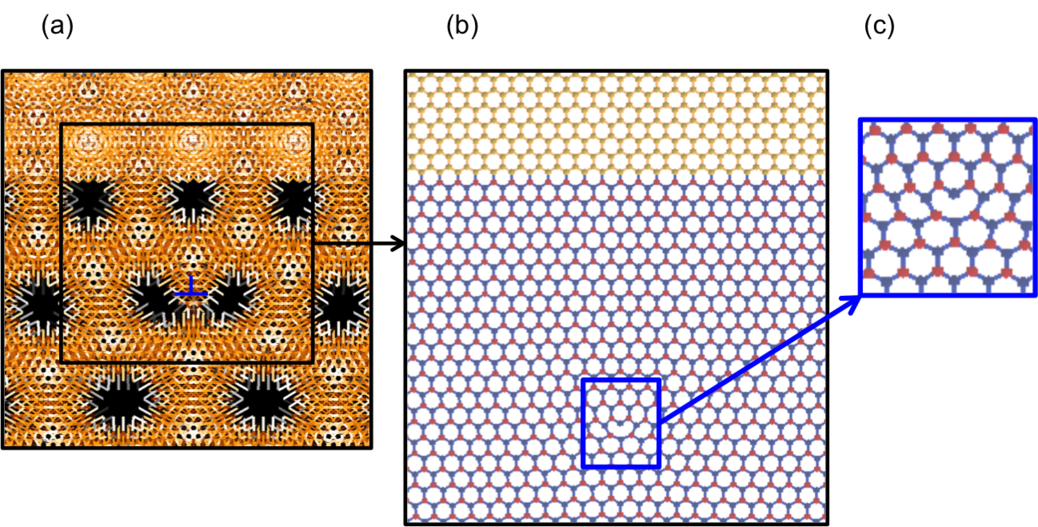}
\caption{\label{fig:interface_atomistic} A moir\'e interface misfit dislocation, such as the one shown in Figure \ref{fig:Moire_disloc}c, is caused by an interface misfit dislocation in one of the constituent layers of the covalent lattices.  (a) The moir\'e dislocation is due to an atomistic dislocation with approximate position shown by the symbol. (b,c) An 8-6 dislocation with precisely the same symmetry is located on the h-BN side, several planes away from the h-BN/graphene interface.}
\end{figure*}

In Figure \ref{fig:Moire_disloc}, experimental and computed moir\'e patterns with interface misfit dislocations are shown -- these patterns may be considered as incoherent moir\'e interfaces.  Here, in Figure \ref{fig:coherent_supp}, a coherent Moir\'e interface is shown.  In this case, an h-BN/graphene lateral heterostructure forms a moir\'e pattern due to misalignment with a Ru(0001) support layer.  But in the part of the interface shown here, there are no misfit dislocations in the underlying covalent lattices, and thus, no moir\'e misfit dislocation.

\begin{figure*}[ht]
\includegraphics[width=5.0in]{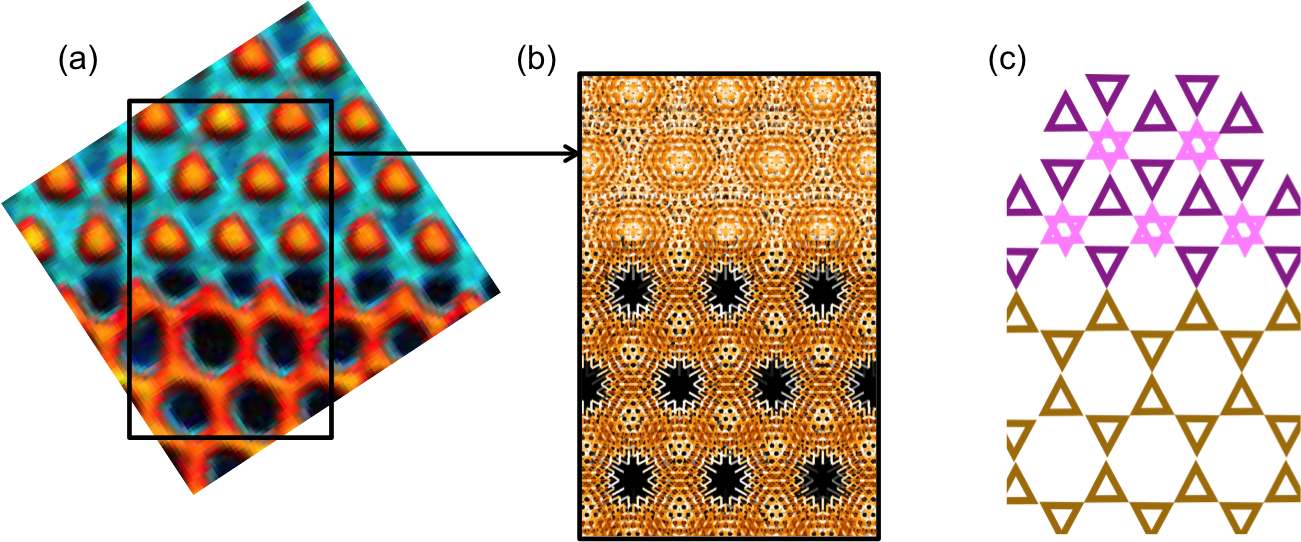}
\caption{\label{fig:coherent_supp} The h-BN/graphene interface, on a Ru(0001) substrate, may produce a coherent moir\'e pattern if there is no underlying strain relieving misfit dislocation.  (a) An experimental image of the coherent interface moir\'e of h-BN/graphene on Ru(0001).\cite{Lu2014} (b) The computed moir\'e pattern shows the inverted contrast on the h-BN and graphene sides, bottom and top, respectively.  (c) The computed moir\'e pattern matches well with the so-called ball-and-stick model, which captures the lines of misalignment between the top layer and substrate.}
\end{figure*}

The example of the linear moir\'e defect, seen when an MoSe$_2$ layer with a mirror-twin boundary interacts with a crystalline graphene substrate, is shown in Figure \ref{fig:Moire_line_defects}(a-d).  As shown in the additional view presented in Figure \ref{fig:Barja_supp}, the experimental defect is remarkably well-matched to the computed moir\'e pattern, though the intensity of the charge density wave is not replicated.  As shown in Figure \ref{fig:Barja_supp}, the moir\'e pattern on both sides of the interface is shifted  between experiment and calculation, highlighting the need of a {\it{screw}}-type antiphase boundary characteristic.
\begin{figure*}[ht!]
\centering
\includegraphics[width=3.5in]{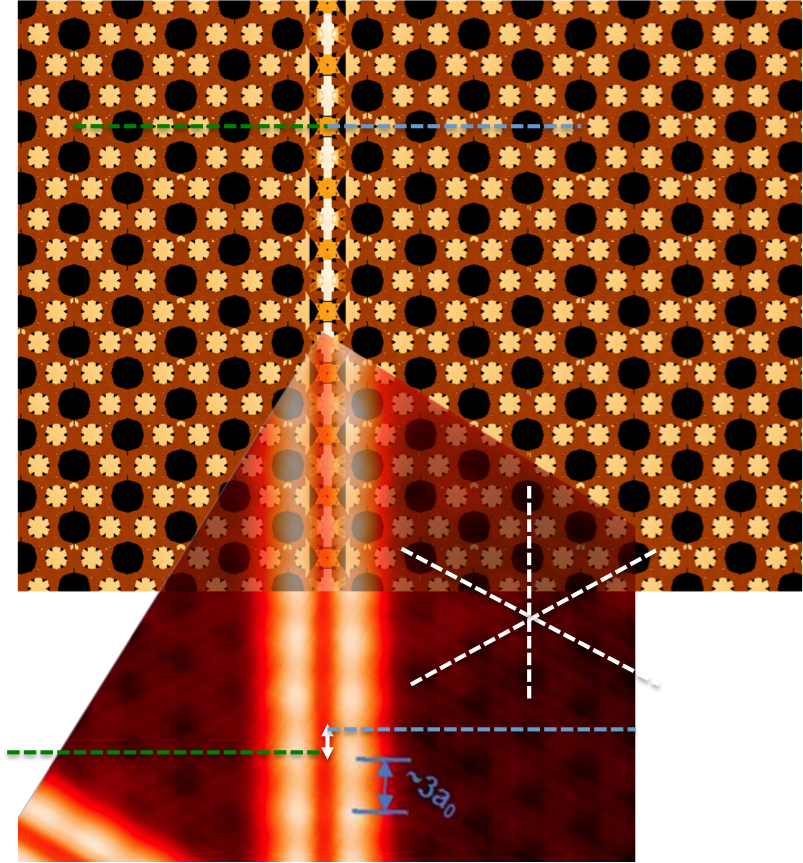}
\caption{\label{fig:Barja_supp} The computed moir\'e pattern for the MoSe$_2$ twin boundary on graphene matches the experimental observation for the same system.{\cite{Barja2016}}  The computed moir\'e symmetry away from the defect matches the experimental symmetry precisely. However, the alignment of both domains is perfect in the simulation while an additional screw shift, which may be attributed to finite size of the triangular domain, is present in the experimental system.} 
\end{figure*}

\begin{figure*}[ht!]
\includegraphics[width=5.5in]{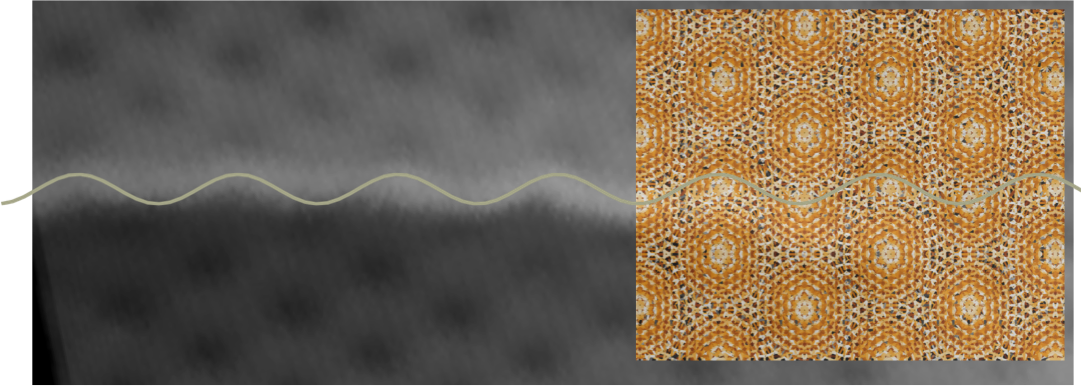}
\caption{\label{fig:Johann_exp_large} The sinusoidal state formed by perfect graphene on an Ir surface step edge, with screw-type APB stacking across the boundary, is presented together with the corresponding Moir\'e simulation.  A sinusoidal curve (grey) is added as a guide to the eye, to facilitate the comparison of the period in both the experiments and the simulation.}  
\end{figure*}

The example of graphene on a screw-type APB on an Ir surface is depicted in Figure \ref{fig:Johann_exp_large}.  A striking sinusoidal state is observed over more than four periods and is compared to its simulated counterpart.  The period corresponds to $\sqrt[]{2}$ times the moir\'e period.  Peaks and troughs of the sinusoid are apart from the two domains, featuring half moir\'e spots on each side.

\begin{figure*}[ht!]
\includegraphics[width=6.5in]{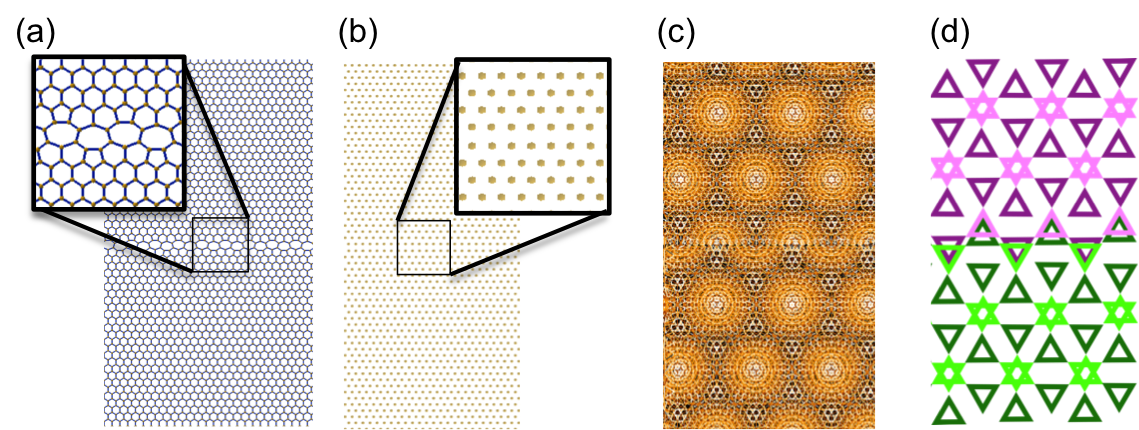}
\caption{\label{fig:zig-zig-defects} A sinusoidal charge density signature can be created by a mixed edge-screw type antiphase boundary in either the 2D material or the substrate.  (a) An antiphase boundary along the zigzag direction in graphene.  (b) An antiphase boundary along the same direction in Ru(0001).  (c) When the first defect is placed on a perfect Ru(0001) substrate, or the second defect is placed under a perfect graphene film, the computed moir\'e pattern contains an identical sinusoidal charge density signature.  (d) In both cases, the ball-and-stick model predicts the feature as a particular overlap of misalignment along the line defect.}  
\end{figure*}

Based on the observations presented in Figure \ref{fig:Moire_line_defects}, one concludes that it is possible to generate a moir\'e defect by exploiting a defect either in the 2D layer or in the metallic support layer.  Thus, one can conceive of more than one system that should exhibit the same type of moir\'e defect, or topological state.  We refer to this as a {\textit{commutativity}} property, because the effect of a defect in one layer will be the same as the effect of the equivalent defect in the other layer.  For example, in Figure \ref{fig:zig-zig-defects}, two {\it{different}} defect combinations lead to the same moir\'e pattern, a sinusoidal topological state along the defect, or anti-phase boundary.  In the one case, a mixed edge- and screw-type antiphase boundary in graphene, along the zigzag direction, is placed over a perfect Ru(0001) substrate.  In the other case, a perfect graphene layer is placed over a screw-type antiphase boundary in the Ru(0001) substrate, with the Ru(0001) APB aligning along the graphene zigzag direction.  We note that the period of this sinusoidal state is equal to the period of the moir\'e lattice, while the period of the sinusoidal state in Figures \ref{fig:Moire_line_defects} and \ref{fig:zig-zig-defects} is equal to $\sqrt[]{2}$ times the moir\'e period, because the orientation of the state here is different.  We also note that the relationship between edge- and screw- character of an APB defect in the hexagonal lattice is different than in the triangular lattice, where it is possible for a boundary to be both pure edge- and pure screw-type due to the symmetry of the lattice.


\section*{References}
\bibliography{VdwMendeley,Moire_bib_extras}


\end{document}